\documentclass[]{aa}
\usepackage{graphicx}
\usepackage[varg]{txfonts}
\usepackage{natbib}
\bibpunct{(}{)}{;}{a}{}{,} % to follow the A&A style
\usepackage[figuresright]{rotating}
\usepackage[a4paper,breaklinks,dvipdfm]{hyperref}

\def\teff{$T\rm_{eff}$}
\def\kms{$\mathrm{km\, s^{-1}}$}

\newcommand{\logg}{\ensuremath{\log {\rm g}}}

\newcommand{\mlp}{\ensuremath{\alpha_{\mathrm{MLT}}}}

\newcommand{\draftflag}{false}

\newcommand{\beq}{\begin{equation}}
\newcommand{\eeq}{\end{equation}}

\begin{document}

\title{X-Shooter GTO: evidence for a population of extremely metal-poor, alpha-poor stars\thanks{Based
on observations obtained at ESO Paranal Observatory,
GTO programme 089.D-0039}
}

\author{
E.~Caffau\thanks{Gliese Fellow}\inst{1,2} \and
P. Bonifacio   \inst{2} \and
P. Fran\c cois \inst{3,2} \and
L. Sbordone    \inst{1,2} \and
M. Spite       \inst{2} \and
L. Monaco      \inst{4} \and
B. Plez        \inst{5} \and
F. Spite       \inst{2} \and
S. Zaggia      \inst{6} \and
H.-G. Ludwig   \inst{1,2} \and
R. Cayrel      \inst{2} \and
P. Molaro      \inst{7} \and
S. Randich     \inst{8} \and
F. Hammer      \inst{2} \and
V. Hill       \inst{9}
}

\institute{ 
Zentrum f\"ur Astronomie der Universit\"at Heidelberg, Landessternwarte, 
K\"onigstuhl 12, 69117 Heidelberg, Germany
\and
GEPI, Observatoire de Paris, CNRS, Univ. Paris Diderot, Place
Jules Janssen, 92195
Meudon, France
\and
UPJV, Universit\'e de Picardie Jules Verne, 33 Rue St Leu, F-80080 Amiens
\and
European Southern Observatory, Casilla 19001, Santiago, Chile
\and
Laboratoire Univers et Particules de Montpellier, LUPM, Universit\'e Montpellier 2, 
CNRS, 34095 Montpellier cedex 5, France
\and
Istituto Nazionale di Astrofisica,
Osservatorio Astronomico di Padova Vicolo dell'Osservatorio 5, 35122 Padova, Italy
\and
Istituto Nazionale di Astrofisica,
Osservatorio Astronomico di Trieste,  Via Tiepolo 11,
I-34143 Trieste, Italy
\and
Istituto Nazionale di Astrofisica,
Osservatorio Astrofisico di Arcetri, Largo E. Fermi 5, 50125 Firenze, Italy
\and
Universit\'e de Nice Sophia Antipolis, CNRS,
Observatoire de la C\^ote d'Azur, Laboratoire Cassiop\'e e, B.P. 4229, 
06304 Nice Cedex 4, France
}
\authorrunning{Caffau et al.}
%\titlerunning{X-Shooter GTO: evidence for a population of extremely metal-poor alpha-poor stars}
\titlerunning{X-Shooter GTO: EMP stars}
\offprints{E.~Caffau}
\date{Received ...; Accepted ...}

\abstract%
%\context
{
The extremely metal-poor stars are the direct descendants
of the first generation stars.
They carry the chemical signature of the pristine Universe
at the time they formed, shortly after the Big Bang.
}
%\aims
{
We aim to derive information about extremely metal-poor stars
from their observed spectra.
}
%\method
{
Four extremely metal-poor stars were selected
from the Sloan Digital Sky Survey (SDSS) and observed
during the guaranteed observing time of X-Shooter.
The X-Shooter spectra were analysed using an automatic
code, MyGIsFOS, which is based on a traditional analysis method. 
It makes use of  
a synthetic grid computed from one-dimensional,
plane-parallel, hydrostatic model atmospheres. 
}
%\results
{
The low metallicity derived from the SDSS spectra is confirmed here. 
Two kinds of stars  are found.
Two stars are confirmed to be extremely metal-poor, with no evidence
of any enhancement in carbon. The two other stars
are strongly enhanced in carbon. We could not derive
iron abundance for one of them, while [Ca/H] is below --4.5.
Two of the stars are members of the rare population of extremely metal-poor
stars low in alpha elements.
}
%\conclusions 
{
}

\maketitle

%%%%%%%%%%%%INTRODUCTION%%%%%%%%%%%%%%%%%%%%%%%%%%%%%%%%%%%

\section{Introduction}

The extremely metal-poor (EMP) stars with a 
metallicity Z lower than about $2.3\times 10^{-3}$ the solar metallicity value,
are probably among the first low-mass stars that formed
in the Universe.
They formed from a material enriched in metals
(all the elements heavier than He) by the matter ejected by
the first generations of massive stars as they
exploded as Type II supernovae.
They were formed shortly after the Big Bang,
at a time when the metal content in the inter-stellar medium
was high enough to allow efficient cooling during
the phase of star formation, so as to form low-mass stars.
The parameters and the chemical composition of these
potentially very old EMP stars can give
us insight into the condition of the primordial gas from which
they formed and into the masses of the Pop.\,III massive stars.
The chemical pattern of these stars, especially the content of carbon,
will tell us something about the relative frequency of the channels for star formation
from a metal-poor gas
in the low-mass regime, i.e. if the cooling of the material occured mainly
through \ion{C}{ii} and \ion{O}{i} \citep{bromm03} or through dust \citep{schneider12}.
In particular, the discovery of SDSS\,J102915+172927 \citep{caffau11}
supports the notion that, at least to form some stars,
the cooling factor must be dust, because C and O abundances in this star are not sufficient
to guarantee the cooling.
For an introduction on metal-poor stars, see \citet{aoki13,sdss_uves,placco11}.

We present here a sample of four EMP stars observed during the
Italian-guaranteed time observation (GTO) of X-Shooter.
The four stars have been selected from SDSS \citep{sdss,yanny09}
to be EMP, and not enhanced in carbon. (For the selection criteria see \citealt{sdss_uves}.)
Two of the stars are EMP, not enhanced in carbon, and one of them
shows low content in $\alpha$-elements (SDSS\,J153346+155701)
when compared to what is found in the majority of metal-poor stars.
The other two stars, SDSS\,J161956+170539 and SDSS\,J174259+253135, happen to be carbon-enhanced stars.
Both these stars are F-type and hot enough, \teff of 6191\,K and 6345\,K respectively, 
to make the G-band not detectable 
at the resolving power of SDSS, and in SDSS\,J161956+170539 the molecular 
band is also hardly visible at the resolution of X-Shooter.
The latter star also is low in $\alpha$-elements.

%%%%%%%%%%%%%%%%%%%%%%%%%%%%%%%%%%%%%%%%%%%%%%%%%%%%%%%%%%%%%%%%%%
\section{Observations and data reduction}

The four stars were observed on the 10 July 2012 with the spectrograph 
X-Shooter \citep{dodorico}
at Kueyen (VLT UT2) during the Italian GTO.
Basic information on the stars are presented in Table\,\ref{allstar},
while the observation log is presented in Table\,\ref{logbook}.
The observation setup and data reduction method are the same
as described in \citet{gto11}. We used the integral field unit 
\citep{IFU} and a $1\times 1$ on-chip binning along the spectral
direction, providing a resolving power R=7\,900 in the UVB arm
and 12000 in the VIS arm. 
In this paper we only report the analysis of the UVB and VIS spectra.

{\small}
\begin{table*}
\caption{\label{allstar}
Coordinates and photometric data.
Optical magnitudes are from the SDSS, infrared magnitudes from 2MASS.}
\tabskip=0pt
\begin{center}
\begin{tabular}{lccrrrrrrrr}
\hline\noalign{\smallskip}
\multicolumn{1}{l}{SDSS ID}& 
\multicolumn{1}{c}{RA}&
\multicolumn{1}{c}{Dec}& 
\multicolumn{1}{c}{l} &
\multicolumn{1}{c}{b} &
\multicolumn{1}{c}{$u$}& 
\multicolumn{1}{c}{$g$}& 
\multicolumn{1}{c}{$r$}&
\multicolumn{1}{c}{$i$}&
\multicolumn{1}{c}{$z$}&
\multicolumn{1}{c}{E(B--V)}\\
 &  J2000.0 & J2000.0 & \multicolumn{1}{c}{deg} & \multicolumn{1}{c}{deg} & [mag] & [mag] & [mag] & [mag] & [mag] & [mag]   \\
\noalign{\smallskip}\hline\noalign{\smallskip}
SDSS\,J144256-001542 & 14\, 42\, 56.37 &$-$00\, 15\, 42.76 &351.88263453 & 51.68619352&$18.78$ & $17.96$ & $17.64$ & $17.52$ & $17.48$ &  $0.042$ \\ 
SDSS\,J153346+155701 & 15\, 33\, 46.28  & +15\, 57\, 01.81 & 24.88837465 & 50.82274117&$17.78$ & $16.90$ & $16.67$ & $16.58$ & $16.56$ &  $0.043$ \\ 
SDSS\,J161956+170539 & 16\, 19\, 56.33  & +17\, 05\, 39.90 & 32.47870056 & 41.01781927&$18.64$ & $17.80$ & $17.53$ & $17.41$ & $17.35$ &  $0.042$ \\ 
SDSS\,J174259+253135 & 17\, 42\, 59.68  & +25\, 31\, 35.90 & 49.85799222 & 25.64468718&$20.06$ & $18.91$ & $18.67$ & $18.55$ & $18.49$ &  $0.065$ \\ 
\noalign{\smallskip}\hline\noalign{\smallskip}
\end{tabular}
\end{center}
\end{table*}

\begin{table*}
\caption{\label{logbook}
Log of the observations.
}
\begin{tabular}{lrrr}
\hline\noalign{\smallskip}
\multicolumn{1}{c}{Star}& 
\multicolumn{1}{c}{date}&
\multicolumn{1}{c}{Exp Time (sec)}& 
\multicolumn{1}{c}{mode}
\\ 
\noalign{\smallskip}\hline\noalign{\smallskip}
SDSS\,J144256-001542 & 2012-07-10 & 1x3690s UVB, 1x3600s VIS, 3x1200s NIR & IFU,Stare; readout: 100k/1pt/hg; binning:1x1  \\ 
SDSS\,J153346+155701 & 2012-07-10 & 1x1290s UVB, 1x1200s VIS, 1x1200s NIR & IFU,Stare; readout: 100k/1pt/hg; binning:1x1  \\ 
SDSS\,J161956+170539 & 2012-07-10 & 1x3090s UVB, 1x3000s VIS, 5x600s  NIR & IFU,Stare; readout: 100k/1pt/hg; binning:1x1  \\ 
SDSS\,J174259+253135 & 2012-07-10 & 2x4800s UVB, 2x4710s VIS, 8x1200s NIT & IFU,Stare; readout: 100k/1pt/hg; binning:1x1  \\ 
\noalign{\smallskip}\hline\noalign{\smallskip}
\multicolumn{4}{l}{
SP STD = spectrophotometric standard;
Ser BLUE = serendipity blue object; RED STD = reddening standard.}
\end{tabular}
\end{table*}

%%%%%%%%%%%%%%%%%%%%%%%%%%%%%%%%%%%%%%%%%%%%%%%%%%%%%%%%%%%%%%%%%%
\section{Analysis}

The analysis was performed with the automatic
code {\tt MyGIsFOS} \citep{sbordone_nic,mygisfos} as in \citet{sdss_uves}.
The main difference is that in this case the
grid of synthetic spectra used by
{\tt MyGIsFOS} was computed using version 12.1.1 of
{\tt turbospectrum} \citep{alvarez_plez,2012ascl.soft05004P}
and OSMARCS 1D LTE model atmospheres 
\citep{G2008}, computed on purpose for this project. 
The grid of plane-parallel models covers the range from 5200\,K to 6800\,K 
with a step of 200\,K in \teff , from 3.5 to 4.5 (c.g.s. units) 
with a step of 0.5 in \logg , and from $-2.5$ to $-4.5$ with a step of 0.5\,dex 
in metallicity. Alpha-element abundances, including oxygen, were varied with 
a step of 0.4 from $-0.4$ to $+0.8$. Model structures were computed for a single 
microturbulence parameter of 1\,km/s. Synthetic spectra were computed at a 
resolution in excess of $\lambda/\Delta\lambda = 400000$, for three microturbulent 
parameters, 0, 1, and 2\,km/s.
The molecular line lists used for calculating the spectra are those listed 
in \citet{G2008}, with the exception of CH from Masseron et al. (in preparation) 
and CN from \citet{hedrosa2013}, both based on improvements in the lists 
of Plez described in \citet{Hilletal2002} and \citet{PC2005}.
For atomic lines we use data from VALD \citep{pisk1995,kupka1999}, 
modified for some of the lines used for the analysis.
The complete list of the atomic data of the lines analysed in this work 
is given in Table\,\ref{lines}.
Since the stars were selected to have turn-off colours, we fixed the gravity 
at \logg = 4.0 (c.g.s units) for this analysis, and 
one single value of microturbulence (1.5\kms) is used (see below).
The grid of synthetic spectra we use in the analysis has been interpolated to this value.

\begin{table}
\caption{\label{lines}
Atomic lines analysed in this work.}
\begin{center}
\begin{tabular}{lrrr}
\hline\noalign{\smallskip}
 Element & $\lambda$   & ${\rm E}_{\rm low}$ & log\,gf \\
         & [nm]        &  [eV]  & \\
\hline\noalign{\smallskip}
%MgI
\ion{Mg}{i}  & 382.9355 & 2.709 & $-0.231$ \\ 
\ion{Mg}{i}  & 383.2299 & 2.712 & $-0.356$ \\ 
\ion{Mg}{i}  & 383.2304 & 2.712 & $+0.021$ \\ 
\ion{Mg}{i}  & 516.7321 & 2.709 & $-0.931$ \\  
\ion{Mg}{i}  & 517.2684 & 2.712 & $-0.450$ \\ 
\ion{Mg}{i}  & 518.3604 & 2.717 & $-0.239$ \\ 
%SiI
\ion{Si}{i}  & 390.5523 & 1.909 & $-0.743$ \\ 
%CaI
\ion{Ca}{i}  & 422.6728 & 0.000 & $+0.265$ \\ 
%CaII
\ion{Ca}{ii} & 849.8023 & 1.692 & $-1.469$ \\ 
\ion{Ca}{ii} & 854.2091 & 1.700 & $-0.514$ \\ 
\ion{Ca}{ii} & 866.2141 & 1.692 & $-0.770$ \\ 
%FeI
\ion{Fe}{i}  & 357.0098 & 0.915 & $+0.153$  \\ 
\ion{Fe}{i}  & 357.0254 & 2.808 & $+0.728$  \\ 
\ion{Fe}{i}  & 374.5561 & 0.087 & $-0.771$  \\ 
\ion{Fe}{i}  & 374.5899 & 0.121 & $-1.335$  \\ 
\ion{Fe}{i}  & 375.8233 & 0.958 & $-0.027$  \\ 
\ion{Fe}{i}  & 381.5840 & 1.485 & $+0.237$  \\ 
\ion{Fe}{i}  & 382.0425 & 0.859 & $+0.119$  \\ 
\ion{Fe}{i}  & 382.4444 & 0.000 & $-1.362$  \\ 
\ion{Fe}{i}  & 385.6371 & 0.052 & $-1.286$  \\ 
\ion{Fe}{i}  & 385.9911 & 0.000 & $-0.710$  \\ 
\ion{Fe}{i}  & 389.5656 & 0.110 & $-1.670$  \\ 
\ion{Fe}{i}  & 393.0297 & 0.087 & $-1.491$  \\ 
\ion{Fe}{i}  & 404.5812 & 1.485 & $+0.280$  \\ 
\ion{Fe}{i}  & 427.1760 & 1.485 & $-0.164$  \\ 
\ion{Fe}{i}  & 427.1153 & 2.449 & $-0.349$  \\ 
\ion{Fe}{i}  & 430.7902 & 1.557 & $-0.073$  \\ 
\ion{Fe}{i}  & 438.3545 & 1.485 & $+0.200$  \\ 
\ion{Fe}{i}  & 440.4750 & 1.557 & $-0.142$  \\ 
\noalign{\smallskip}\hline\noalign{\smallskip}
\end{tabular}
\end{center}                   
\end{table}

As in our previous paper \citep{gto11},
we derived the effective temperature from the SDSS $\left(g-z\right)$ colour,
by means of the calibration presented in \citet{ludwig08}.
The reddening adopted was the one provided by the SDSS data base, which is
from the \citet{schlegel} maps.
We also derived \teff\ by fitting the wings of H$\alpha$,
with synthetic
profiles computed using a modified version of the
{\tt BALMER} code\footnote{The original version is available on-line
at \url{http://kurucz.harvard.edu/}}, which uses the theory of
\citet{barklem00,barklem00b} for self-broadening and Stark broadening from  
\citet{stehle99}. The model atmospheres for the H$\alpha$ calculations were computed
using version 9 of the ATLAS code \citep{kurucz93,kurucz05,sbordone04,sbordone05} 
and \mlp = 0.5, as recommended by 
\citet{fuhrmann} and \citet{vantveer}.

The fit on  the H$\alpha$ wings 
in extremely metal-poor stars 
is sensitive to the gravity \citep{sbordone}.
A change of $^{+}_{-}0.5$\,dex in gravity for a star at 6350\,K 
would change the temperature by $^{-228}_{+87}$\,K.
For two stars (SDSS\,J144256-001542 and SDSS\,J153346+155701),
the temperatures derived with the two methods do not agree.
SDSS\,J144256-001542 implies a lower temperature from H$\alpha$ wings fit
(\teff = 6030\,K and 5850\,K with \logg = 3.5 and 4.0, respectively, 
to be compared to \teff = 6161\,K from $g-z$ colour).
The reddening from the Schlegel maps \citep{schlegel}
is very low (E(B--V)$\sim 0.04$) and  could be slightly overestimated.
Assuming no reddening, which is consistent with the small equivalent
width of the \ion{Ca}{ii}-K
interstellar component, \teff\ from photometry (5914\,K) is in good
agreement with the one derived from the H$\alpha$ wings fit.
We decided for this star to keep \teff\ derived from H$\alpha$
and \logg = 4.0.
Also for SDSS\,J153346+155701, the \teff\ from the $(g-z)_0$ colour 
obtained assuming the reddening provided by the
\citet{schlegel} maps is 124\,K higher than what is implied
by the H$\alpha$. In this case the interstellar \ion{Ca}{ii} K line is
slightly stronger than in SDSS\,J144256-001542, but still
weak. We therefore prefer the temperature derived from the 
H$\alpha$ wings fit (\teff = 6375\,K), which is consistent 
with a reddening that is non-zero, but lower than estimated
from the \citet{schlegel} maps (E(B--V) = 0.01 rather than 0.04).

The microturbulence cannot be derived from X-Shooter spectra
because weak lines are not detectable at low (R=7\,900) resolving power.
We attempted to derive a calibration with \teff\ and \logg,
which is relevant for the extremely metal-poor stars, 
assembling all the stars in the samples of \citet{sbordone} and
\citet{sdss_uves} for a total of 45 stars (the
different spectra of the same stars in \citet{sdss_uves} counting
as different stars). For the stars in the present sample, the
calibration provides values of the microturbulence between 1.43\,\kms
and 1.56\,\kms. Considering that the r.m.s of the calibration is 0.3\,\kms,
we decided to adopt 1.5\,\kms for all our stars.

All four target stars turned out to be extremely metal-poor, and
two of them happened to also be carbon-enhanced
(CEMP stars SDSS\,J161956+170539, SDSS\,J174259+253135), although the carbon enhancement
was not at all apparent in the SDSS spectrum.
The resolution of X-Shooter allowed us to detect the C abundance from the G-band,
but higher resolution observations are desirable.
For SDSS\,J174259+253135, an observing programme with UVES@VLT is on-going.
Both these stars fall in the {\em lower} carbon plateau discussed by \citet{Spite13}
in their figure\,14.

%%% FIGURE %%%%%%%%%%%%%%%%
\begin{figure}
\begin{center}
\resizebox{\hsize}{!}{\includegraphics[draft = \draftflag,clip=true]
{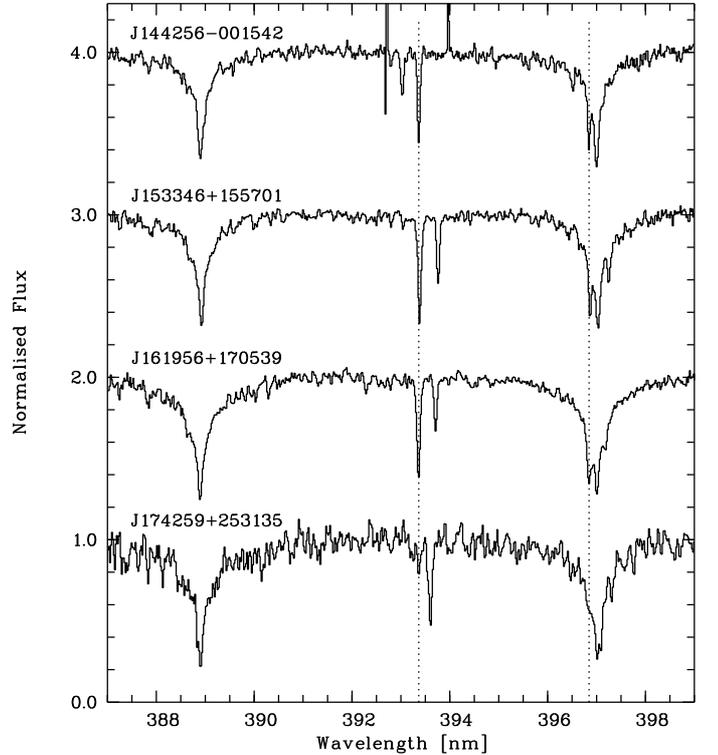}}
\end{center}
\caption[]{The range of the \ion{Ca}{ii}-K and -H of the four stars.
The extreme Ca deficiency of SDSS\,J174259+253135 is clearly visible 
in comparison to the three other stars of the sample.
The dotted lines mark the positions of the stellar \ion{Ca}{ii} lines 
to help to distinguish them from the interstellar lines.
}
\label{plotcak}
\end{figure}
%%% FIGURE %%%%%%%%%%%%%%%%

The results of our analysis are summarised in   Tables\,\ref{analysis}
and \ref{abbondanze}.
The solar abundances are from \citet{lodders09} for Mg, Si, and Ca,
and from \citet{abbosun} for Fe and C.
The table also reports
our estimate of [Fe/H] obtained from the SDSS spectra.
This is {\em not} the estimate provided by the SDSS
pipeline \citep{carlos_sdss}, but the one obtained by our
analysis of the SDSS spectra. 
This estimate has already been mentioned in our previous papers
\citep{bonifacio11,sdss_uves,gto11}. 
It has been obtained from our code for automatic
analysis Abbo \citep{bonifacio03}. The effective temperature is
determined from the $(g-z)_0$ colour,
the surface gravity is fixed to \logg =4.0, and microturbulence
is set to 1\, kms$^{-1}$. The [Fe/H] and 
[$\alpha$/Fe] are determined by a $\chi^2$ fit to ten prominent
spectral features, which are measurable at the resolution
of SDSS spectra. These include \ion{Ca}{ii} K line and infrared
triplet, the \ion{Ca}{I} 422\,nm line, the \ion{Mg}{I}b triplet, 
the G-band, and several features
dominated by \ion{Fe}{i} lines. 
At very low metallicities most features have to be discarded since they are too weak 
or too noisy, so that the estimate is essentially based
on the \ion{Ca}{ii} K line, and [$\alpha$/Fe] 
is undetermined. The grid of synthetic spectra employed
assumes [Ca/Fe]=+0.4. 

\begin{figure}
\begin{center}
\resizebox{\hsize}{!}{\includegraphics[draft = \draftflag,clip=true]
{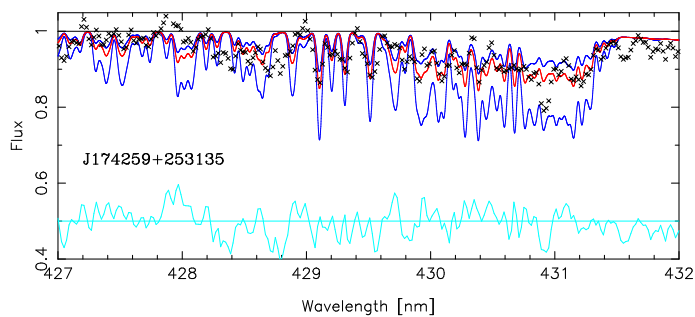}}
\end{center}
\caption[]{G-band of the observed spectrum (black crosses) of the star SDSS\,J174259+253135. 
Super-imposed is the best fit with A(C)=7.2 (solid red), and synthetic profiles with A(C) 7.4 and 7.8 (solid blue),
on either side of the best fit.
In the lower part of the figure, the difference between the best fit and the observed spectrum (solid cyan) is shown and
it has been displaced vertically by 0.5 for display purposes.
}
\label{gband174259}
\end{figure}

Star SDSS\,J174259+253135
is the faintest in the sample, and the spectrum shows the lowest S/N.
In Fig.\,\ref{plotcak} the range of the \ion{Ca}{ii}-K and -H lines
is shown for all four stars, and the low S/N of SDSS\,J174259+253135 is evident. 
It is also evident that this star is the most Ca-poor
star of the sample.
We detect in the X-Shooter spectrum only the G-band
(see Fig.\,\ref{plot_metcomp}) and the \ion{Ca}{ii}-K line,
which is, however, strongly contaminated by interstellar absorption.
This is not surprising considering its low Galactic
latitude. We derive an upper limit on the Ca abundance of
[Ca/H]$\le -4.5$.
From the G-band we deduce that the 
star is strongly enhanced in carbon (A(C)$\sim 7.4$, in Fig.\,\ref{gband174259}
the best fit is shown).
To derive an upper limit on the iron
abundance in this star, we applied the Cayrel formula
\citep{cayrel88} to the two strongest \ion{Fe}{i}
lines expected in our spectrum: 
382.0\,nm and 385.9\,nm.
Applying a $2\sigma$ criterion, we estimated the Fe abundance
needed to produce an equivalent width twice the error
provided by the Cayrel formula. This implies 
[Fe/H]$<-3.8$.  

%%% FIGURE %%%%%%%%%%%%%%%%
%%% FIGURE %%%%%%%%%%%%%%%%

\begin{figure}
\begin{center}
\resizebox{\hsize}{!}{\includegraphics[draft = \draftflag,clip=true]
{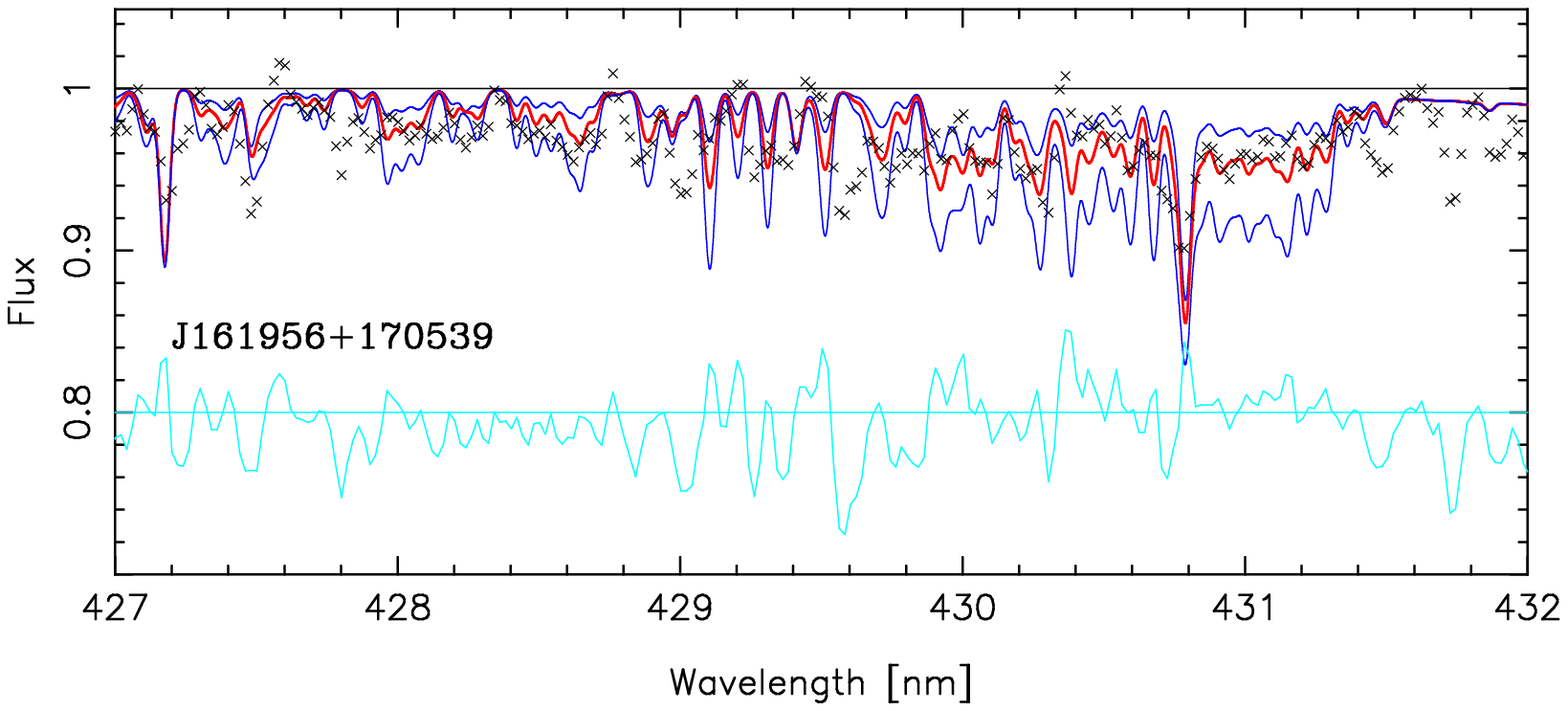}}
\end{center}
\caption[]{G-band of the observed spectrum (black crosses) of the star SDSS\,J161956+150539. 
Super-imposed is the best fit with A(C)=6.8 (solid red), and synthetic profiles with A(C) of 6.5 and 7.1 (solid blue),
on either side of the best fit.
In the lower part of the figure, the difference between the best fit and the observed spectrum (solid cyan) is shown and
it has been displaced vertically by 0.8 for display purposes.
}
\label{gband161956}
\end{figure}

\begin{figure}
\begin{center}
\resizebox{\hsize}{!}{\includegraphics[draft = \draftflag,clip=true]{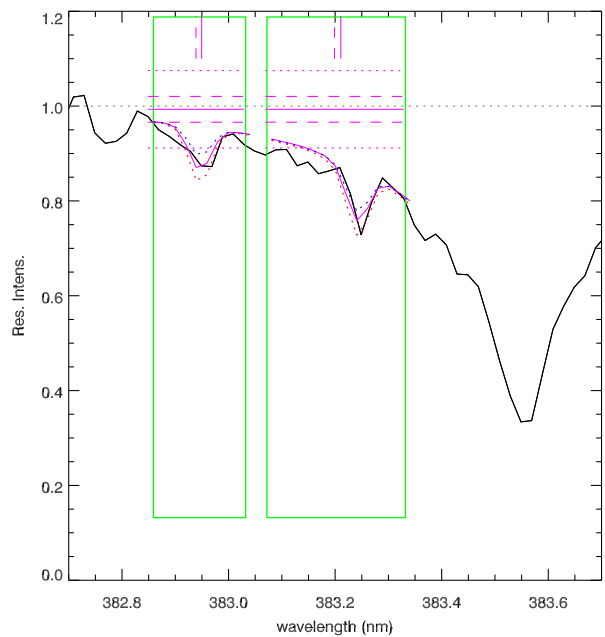}}
\end{center}
\caption[]{The observed spectrum of two \ion{Mg}{i} lines (solid black) in SDSS\,J161956+170539
with the best fit (solid pink) and the best fitting overall (dashed blue) super-imposed.
Two synthetic spectra (dotted pink and blue, respectively) are overplotted
with Mg abundance $\pm 0.3$\,dex with respect to the best fit.
}
\label{mgsdssj161956}
\end{figure}

We also analysed the G-band 
for SDSS\,J161956+170539 (see Fig.\,\ref{gband161956}), and derived A(C)=7.2.
This CEMP star seems peculiar, because it is poor in $\alpha$-elements.
Its oxygen abundance cannot be derived. We derived an upper limit
of [O/H]$< -1$ from the \ion{O}{i} triplet lines at 777\,nm that
is not significant.
Its abundance of Mg is based on four lines, with a low line-to-line scatter (0.11\,dex).
The [Mg/Fe] we derive is +0.04, and in Fig.\,\ref{mgsdssj161956} the best fit of two
\ion{Mg}{i} lines is shown.
Changes in the stellar parameters do not alter the picture for Mg, maintaining
the star poor in Mg.
A change in \teff\ of $\pm 200$\,K alters [Mg/Fe] by $^{+0.20}_{-0.23}$\,dex,
and a change in \logg\ of $\pm 0.5$\,dex changes [Mg/Fe] by $^{-0.09}_{+0.06}$\,dex.
In all these cases the line-to-line scatter increases to 0.22, up to 0.38\,dex.

\begin{table*}
\caption{\label{analysis}
The stellar parameters and main results.
}
\begin{center}
\setlength{\tabcolsep}{2pt}
\begin{tabular}{lrrrrrrrr}
\hline\noalign{\smallskip}
Star                 & ${\rm T_{\rm eff}}$ & \logg & $\xi$ &S/N & [Fe/H]$_{\rm SDSS}$ & [Fe/H] & [$\alpha$/H] & A(C) \\
                     & K & [c.g.s.] & \kms & @\,400\,nm &    \\
\hline\noalign{\smallskip}
SDSS\,J144256-001542 & 5850 & 4.0 & 1.5 &  40 & $-3.36$ & $-4.09\pm 0.21$ & $-3.81$ &     \\
SDSS\,J153346+155701 & 6375 & 4.0 & 1.5 &  47 & $-3.19$ & $-3.34\pm 0.26$ & $-3.28$ &     \\
SDSS\,J161956+170539 & 6191 & 4.0 & 1.5 &  46 & $-3.39$ & $-3.57\pm 0.25$ & $-3.61$ & 7.2 \\
SDSS\,J174259+253135 & 6345 & 4.0 & 1.5 &  20 & $-6.06$ & $<-3.8$         & $ - $   & 7.4 \\
\noalign{\smallskip}\hline\noalign{\smallskip}
\multicolumn{9}{l}{
S/N @\,400\,nm is per extracted rebinned pixel, which is of 0.02\,nm.}
\end{tabular}
\end{center}
\end{table*}

\begin{table*}
\setlength{\tabcolsep}{2pt}
\caption{Abundances \label{abbondanze}}
\centering
\begin{tabular}{lcccccccccccccr}
\hline\hline
Star        &[Fe/H]& [Mg/H] & $\sigma$ & N &  [Si/H]&  N &$\sigma$ & [\ion{Ca}{i}/H]  & $\sigma$ & N &
  [\ion{Ca}{ii}/H] & $\sigma$ & N \\ 
\hline
\object{SDSS\,J144256+253135} & $-4.09$ & $-3.82$ & 0.18 & 1 &       & &      & $-3.80$ & 0.19 & 1 & $-3.69$ & 0.08 & 2 \\
\object{SDSS\,J153346+155701} & $-3.34$ & $-3.28$ & 0.13 & 3 &       & &      & $-3.26$ & 0.19 & 1 & $-2.40$&  0.24 & 3  \\ 
\object{SDSS\,J161956+170539} & $-3.57$ & $-3.53$ & 0.11 & 4 &$-3.88$&1& 0.19 & $-3.92$ & 0.19 & 1 & $-3.20$&  0.23 & 2 \\   
\hline
\multicolumn{15}{l}{ The $\sigma$ represents the line-to-line scatter when more than one line is measured.
It is derived from}\\
\multicolumn{15}{l}{ Monte Carlo simulations when only one line is measured (see
Sbordone et al. 2013)}
%----------- ------------- ----------------------------------------------------------------------------------------------
\end{tabular}
\end{table*}

%%% FIGURE %%%%%%%%%%%%%%%%
%%% FIGURE %%%%%%%%%%%%%%%%

%%% FIGURE %%%%%%%%%%%%%%%%
%%% FIGURE %%%%%%%%%%%%%%%%

Two stars (SDSS\,J153346+155701 and SDSS\,J161956+170539) show a substantial difference
in the Ca abundance derived from \ion{Ca}{i} and \ion{Ca}{ii} lines.
This can be explained by the fact that
the region of the \ion{Ca}{ii} triplet at 850\,nm is contaminated by telluric absorption, 
and the sky subtraction is difficult. The effects of departures from local thermodynamical
equilibrium acts in opposite ways on the \ion{Ca}{i} and \ion{Ca}{ii} lines
\citep{EC12}; these effects will be computed in a dedicated paper.

None of the stars shows an identifiable Li feature at 670.7\,nm, but the
upper limit that we can derive, owing to the resolution of X-Shooter and the S/N ratio 
of these spectra, is not conclusive.
Only SDSS\,J174259+253135 shows a feature at the wavelength where the Li doublet is expected.
The high-resolution UVES spectra will allow us to derive a conclusive answer for Li.

\section{Kinematics}

The radial velocities measured from our X-Shooter spectra
are within 15\,\kms \ of the radial velocities measured
by SDSS. This is within the expected combined error
of the two measurements and thus does not support the existence of 
radial velocity variations for any of the stars. 
All the objects of the present sample have an extreme radial velocity
that makes them compatible with a Halo population of stars. 
For each star we compared the observed radial velocity with a
Besan\c{c}on \citep{Robin} simulation with kinematics of the field in
the direction of each star.

The SDSS radial velocity of SDSS J144256-001542 is 225\,km/s. The
comparison with the Besan\c{c}on simulation shows that the radial velocity is at the
upper extreme of the galactic halo velocity dispersion,
i.e. $\simeq2.8\times$ higher than velocity distribution of the
stars within 0.1\,mag from the stars' colour and magnitude.

Star SDSS J153346+155701 has a radial velocity of $-309$\,km/s, which
makes it an extreme halo star also considering the position in the
colour-magnitude diagram of this star. In fact, in the (g-i) colour, this
star is 0.3\,mag bluer than the bulk of the  turn-off population in field. In
the corresponding Besan\c{c}on simulation, there are very few stars to be
compared with.

Also, the last two stars J161956+170539 and the CEMP star
J174259+253135 are possible members of the halo population. Both
radial velocities at $-324\pm 4$\,km/s for the first and $-221\pm 11$\,km/s for the
second are compatible with the expected velocity distribution in the
respective fields of the Besan\c{c}on simulations.

%%%%%%%%%%%%%%%%%%%%%%%%%%%%%%%%%%%%%%%%%%%%%%%%%%%%%%%%%%%%%%%%%%%%%%%%%%%%%%%%%%%
\section{Discussion}

This last run of our X-Shooter GTO has confirmed the high
reliability with which we can select EMP stars from the SDSS
spectra. All four stars are confirmed EMPs from our spectra.
It is interesting to note that, up to very recently, the binary system
CS\,22876-32 \citep{molaro90,norris00,jonay} has provided the only 
unevolved stars at [Fe/H] well below --3.5. Our own studies 
\citep{caffau11,gto11,EC12,sdss_uves,Spite13} have found    
five more stars in this metallicity regime, and this paper
adds three more stars to the sample.

\begin{figure}
\begin{center}
\resizebox{\hsize}{!}{\includegraphics[draft = \draftflag,clip=true]
{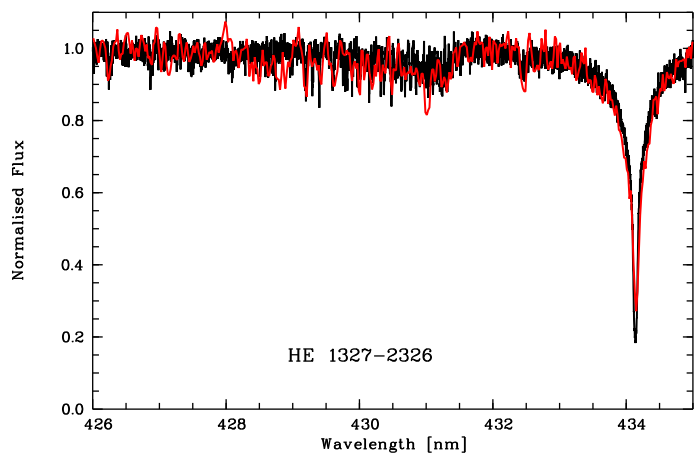}}
\end{center}
\caption[]{The observed spectrum (black crosses) of the star SDSS\,J174259+253135
in the range of the G-band, compared to the CEMP from \citet{frebel} (solid red). 
}
\label{plot_metcomp}
\end{figure}

\begin{figure}
\begin{center}
\resizebox{\hsize}{!}{\includegraphics[draft = \draftflag,clip=true]
{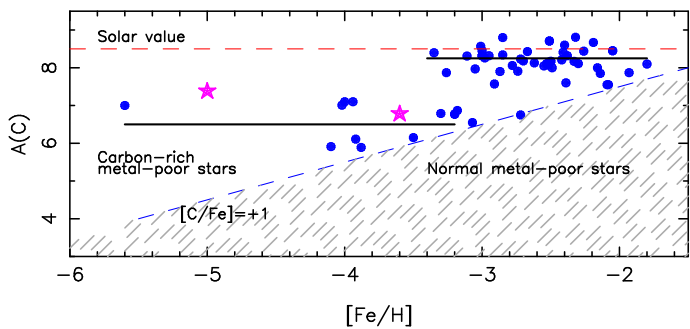}}
\end{center}
\caption[]{The carbon abundance, A(C), versus [Fe/H] for the two CEMP stars (pink stars)
compared to the sample of \citet{Spite13} (blue solid circles).
The dashed (red) horizontal line is the solar photospheric carbon abundance.
The solid horizontal (black) lines are the two carbon plateaux for CEMP stars,
tentatively identified by \citet{Spite13}.
The dashed (blue) tilted line, corresponding to [C/Fe]=+1, separates CEMP stars
from C-normal stars.}
\label{plateau}
\end{figure}

Star SDSS\,J174259+253135 is clearly the most Fe-poor star of
the present sample. Its prominent G-band declares it a CEMP star.
A direct comparison of its spectrum with that of
HE\,1327-2326 \citep[${\relax\rm [Fe/H]} \sim -6$][]{frebel,Frebel08} 
shows that the G-bands of the two stars are quite similar,
as shown in Fig.\,\ref{plot_metcomp},
given that their effective temperatures are the same to  within 200\,K,
the carbon abundances must be roughly the same.
It is suggestive that both stars appear on the {\em lower} carbon
plateau highlighted by \citet{Spite13} and presented here in Fig.\,\ref{plateau}.
Better understanding of the chemical composition of this star
will follow when a high-resolution spectrum is available.

%%% FIGURE %%%%%%%%%%%%%%%%
%%% FIGURE %%%%%%%%%%%%%%%%

All three other stars appear to be more metal-poor
than our estimates
from the SDSS spectra. 
\citet{gto11} compared the estimates
from the SDSS spectra with the ones based on 
X-Shooter or UVES spectra in their figure\,5. 
The only star in that plot 
that has a metallicity below $-3.5$, for which 
the higher resolution spectrum provided an abundance
{\em lower} than estimated from the SDSS spectrum,
was SDSS\,J102915+172927. Although in \citet{gto11}
it was argued that below this metallicity the estimate derived
from SDSS spectra had a large error (about 1\,dex), but was unbiased. 
The present results support the argument in \citet{gto11} that the 
metallicity estimate is indeed unbiased and that  
the SDSS spectra {\em do not} systematically
underestimate the metallicity for these stars.

%%% FIGURE %%%%%%%%%%%%%%%%
%%% FIGURE %%%%%%%%%%%%%%%%

\begin{figure}
\begin{center}
\resizebox{\hsize}{!}{\includegraphics[clip=true]
{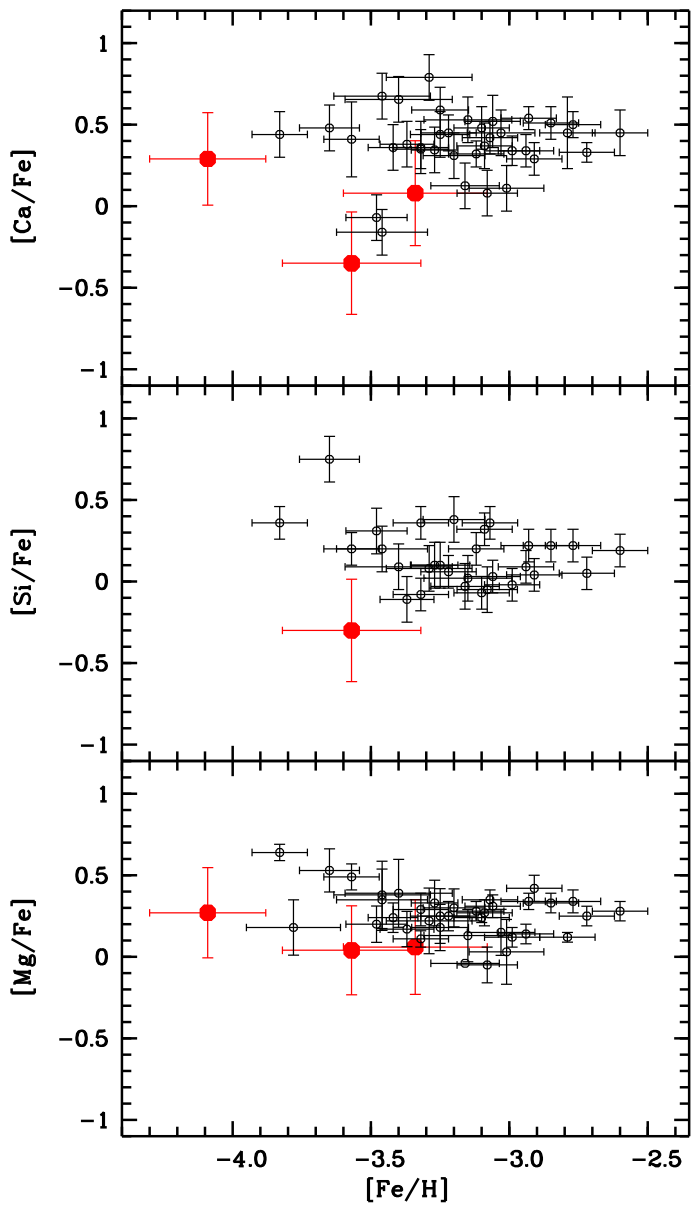}}
\end{center}
\caption[]{The $\alpha$ elements in the programme stars (filled red symbols)
compared with those in the samples of Bonifacio et al. (2009) and (2012) 
(black symbols). 
The error bars on [X/Fe] has been computed by adding under quadrature
the errors on [X/H] and [Fe/H], given in Tables\,\ref{analysis} and \ref{abbondanze}.
}
\label{plalpha}
\end{figure}

In Fig.\,\ref{plalpha}, we compare $\alpha$-elements versus [Fe/H] from this analysis
to the samples of EMP stars from Bonifacio et al. (2009) and (2012).
The abundances of Ca are based on \ion{Ca}{i} lines.
The silicon abundances  are based on the
only available line for metal-poor TO stars: \ion{Si}{i} 390.6\,nm
\citep{bonifacio09}. It is not surprising that it can
be measured in one star, but is too weak so has to be discarded in all the other cases.
Of the three stars for which we can measure the abundance of Fe and
$\alpha$ elements, two stars do not present a significant enhancement
in the $\alpha$-to-iron ratio.
Star SDSS\,J144256+253135 shows an enhancement in the $\alpha$-elements,
Mg and Ca, as expected for EMP stars.
In contrast, SDSS\,J153346+155701 only shows a tiny enhancement in [Mg/Fe] and 
[Ca/Fe], and SDSS\,J161956+170539 is depleted in $\alpha$-elements.
Such  low-$\alpha$ stars, in this very low metallicity regime, are becoming 
evident now that the sizes of the samples increase.
A few such stars at slightly higher metallicities
are already known: BD+~80$^\circ$~245
 \citep{carney} with $[{\rm Fe/H}]=-1.8$,
CS 22873-139 \citep{spite00} with $[{\rm Fe/H}]=-3.4$, and
SDSS\,J135046+134651 \citep{bonifacio11} with $[{\rm Fe/H}]=-2.3$.
Since, the numbers of the analysis of these stars increase, it will be possible to
find out whether they constitute a distinct stellar population
or if they are somehow peculiar.
Such stars have also been found in metal-poor dwarf spheroidal galaxies
(see \citealt{tafelmeyer10,Frebel10,starkenburg13,venn12}, and for a review see
\citealt{hill12}). 

%%% FIGURE %%%%%%%%%%%%%%%%
%%% FIGURE %%%%%%%%%%%%%%%%

\citet{nissen10} in their sample of halo stars highlight
two populations, a high-$\alpha$ population, showing a constant
value of [$\alpha$/Fe] vs. [Fe/H]; and a low-$\alpha$ population,
with [$\alpha$/Fe] vs. [Fe/H] decreasing with increasing metallicity.
They suggest that the high-$\alpha$ stars formed from gas
enriched by Type\,II supernovae in a region with a high star formation rate.
The low-$\alpha$ population should be of stars formed in region
where the chemical evolution is slow and the Type\,I supernovae had
time to contribute in iron, inducing the negative trend of [$\alpha$/Fe]
increasing metallicity.
Our low [$\alpha$/Fe] stars are too metal-poor and too 
depleted in $\alpha$-elements
to be part of this population. The question remains
whether these stars and the Nissen\& Schuster low-$\alpha$  stars
are somehow related.
If the EMP low-$\alpha$ stars are interpreted as 
the relic of a low star formation region of the Galaxy
(or former satellite galaxy), then they could be several
Gyr younger than the rest of the halo.
A suggestive example of this kind of occurrence is
shown by \citet{RJ12}, who simulate the evolution of a 
dwarf galaxy of $3\times 10^8 M_\odot$. Their figure 6 shows
the [Fe/H] vs. [Mg/Fe] diagram for this simulated galaxy, which displays
a rare, but definite, population of low-$\alpha$ stars at metallicities 
centred on --2.0. The galaxy shows a strong starburst at the 
beginning of its life and a weak secondary starbust about 2 Gyr after.
The low-$\alpha$ population is formed in this secondary peak, when
the Type Ia SNe had time to enrich the gas in Fe, at constant Mg abundance.
Our low-$\alpha$ stars could have been formed
in dwarf galaxies of this kind.
In principle, one 
possibility of forming  EMP stars with low oxygen-to-iron ratios is
to require  an environment dominated by Type\,II supernovae
of relatively low-mass ($< 20 {\rm M}_{\odot}$).
The yields of zero-metallicity stars computed by \citet{LC}
for their models of 13 and 15 M\sun\ imply low [O/Fe]
ratios (0.1 -- 0.2); however, the corresponding [Mg/Fe] and [Ca/Fe]
ratios show no clear dependence on mass and in any case
[Ca/Fe] $< 0$ is not seen for any of their models.
Thus this occurrence, by itself, cannot explain our observations.

\citet{cayrel04} have found a very small scatter in the $\alpha$ elements in their
sample of extremely metal-poor stars. They conclude
that the gas from which the
stars formed was well mixed. 
With these EMP stars with low-$\alpha$ to iron
ratios, we are induced to conclude that the ``well mixed halo'' 
does not hold for our sample. 
It is possible that we are 
sampling a more distant population than the
giants of \citet{cayrel04}\footnote{The distances of giant stars are uncertain
to the extent that their surface gravities are poorly constrained.
The giants in \citet{cayrel04} should be 2.5 to 4.5 magnitudes
brighter than corresponding TO stars. A typical giant
of the \citet{cayrel04} sample has V=13.5 and is at comparable
distances to TO stars of magnitudes 16 to 18. 
The uncertainties in the distances of both giants and dwarfs are 
so large that it is not possible to decide if the two
samples are at comparable distances or if one is more distant
than the other, 
but it is also possible that it is the lower mean
metallicity of our sample 
with respect to that of \citet{cayrel04}
that is the main factor behind this difference. 
}

%%%%%%%%%%%%%%%%%%%%%%%%%%%%%%%%%%%%%%%%%%%%%%%%%%%%%%%%%%%%%%%%%%%%%%%%%%%%%%%%%%%
\section{Conclusions}

This last X-Shooter GTO run allowed us to discover four new
extremely metal-poor stars, some of which display some
outstanding characteristics.
The CEMP star SDSS\,J174259+253135 shows no detectable metallic
lines other than a weak \ion{Ca}{ii} K line, this
makes it a good candidate to be one of the most Fe-poor
stars known.
Star SDSS\,J161956+170539 shows a peculiar chemical pattern,
where the carbon enhancement is accompanied by a low value of
measured $\alpha$-elements, making it a unique object.
It is remarkable that {\em both} these CEMP stars 
lie on the {\em lower} carbon plateau \citep{Spite13}, 
helping to confirm its reality.
The two carbon-normal stars seem to belong to two distinct
populations. 
SDSS\,J144256+253135 is a member 
of the usual $\alpha$-enhanced halo population,
while 
SDSS\,J153346+155701 belongs to a rare 
$\alpha$-poor population at extremely low metallicity, 
possibly the EMP counterpart of the Nissen \& Schuster 
low-$\alpha$ population.
These findings stress the importance of both
increasing the sample sizes and the surveyed volume, reaching
for more distant stars.

We may summarise the results of our X-Shooter GTO programme,  now finished. This
covered 2.5 nights of observations (1.5 nights of Italian and 1 night of
French GTO) and 0.75 nights were lost due to bad weather.
We could observe twelve stars in the 
$g$ magnitude range 17 to 19, half of which have [Fe/H]$<-3.5$.
In the sample, SDSS\,J102915+172927 happens to be the most metal-poor star
known to date \citep{caffau11}; SDSS\,J174259+253135 shows no evident
feature except the \ion{Ca}{ii}-K line and the G-band; three stars are EMP
and low in $\alpha$-elements (SDSS\,J082511+163459 \citealt{gto11}, 
SDSS\,J153346+155701, and SDSS\,J161956+170539); SDSS\,J135516+001319 is a 
metal-poor star that is very low in $\alpha$-elements \citep{bonifacio11}.

%%%%%%%%%%%%%%%%%%%%%%%%%%%%%%%%%%%%%%%%%%%%%%%%%%%%%%%%%%%%%%%%%%%

\begin{acknowledgements}
We are grateful to Pascale Jablonka for her
useful comments on our manuscript.
EC, LS, and HGL acknowledge financial support
by the Sonderforschungsbereich SFB881 ``The Milky Way
System'' (subprojects A4 and A5) of the German Research Foundation
(DFG).
PB, PF, MS, FS, and RC acknowledge support from the Programme National
de Cosmologie et Galaxies (PNCG) of the Institut National de Sciences
de l'Univers of CNRS.
This work has made use of the VALD database, operated at Uppsala University, 
the Institute of Astronomy RAS in Moscow, and the University of Vienna.
\end{acknowledgements}

%%%%%%%%%%%% APPENDIX %%%%%%%%%%%%%%%%%%%%%%%%%%%%

%
%  A&A article: Metal-poor stars
%
%
%%%%%%%%%%%%%%%%%%%%%%%%%%%%%%%%%%%%%%%%%%%%%%%%%%%%%%%%%%%%%%%%%%%%%%%%%%%
%\appendix
%%%%%%%%%%%%%%%%%% END APPENDIX %%%%%%%%%%%%%%%%%%%%%%%%%%%%%%%

\bibliographystyle{aa}

\end{document}